# 3-D Radiation Mapping in Real-Time with the Localization and Mapping Platform LAMP from Unmanned Aerial Systems and Man-Portable Configurations

R. Pavlovsky, A. Haefner, T.H. Joshi, V. Negut, K. McManus, E. Suzuki, R. Barnowski, K. Vetter

ABSTRACT
Real-time, meter-resolution gamma-ray mapping is relevant in the detection and mapping of radiological materials, and for applications ranging from nuclear decommissioning, waste management, and environmental remediation to homeland security, emergency response, and international safeguards. We present the Localization and Mapping Platform (LAMP) as a modular, contextual and radiation detector sensor suite, which performs gamma-ray mapping in three dimensions (3-D) and in real time, onboard an unmanned aerial vehicle (UAV) or in a man-portable configuration. The deployment of an unmanned aerial system (UAS) for gamma-ray mapping can be advantageous, as the UAS provides a means of measuring large areas efficiently and improving accessibility to some environments, such as multi-story structures. In addition, it is possible to increase measurement robustness through autonomous navigation, and to reduce radiation exposure to users as a result of the remote measurement. LAMP enables meter-resolution gamma-ray mapping through Scene Data Fusion (SDF) [1], a capability that fuses radiation and scene data via voxelized 3-D Maximum Likelihood Expectation Maximization (MLEM) to produce 3-D maps of radioactive source distributions in real-time. Results are computed onboard LAMP while it is flying on the UAV and streamed from the system to the user, who can view the 3-D map on a tablet in real-time. Modularity of the LAMP system enables customization of sensor combinations tailored to specific use cases. Information about the scene, i.e. the surrounding environment, is collected using contextual sensors, including a Velodyne Puck LiDAR sensor [8], while gamma-ray data is collected using four Kromek Sigma50 CsI detectors [9]. We present results that demonstrate the SDF concept, including a set of UAS flights where a $^{133}$Ba source is localized at a test site in Berkeley, CA and a handheld measurement in Fukushima Prefecture, Japan where the distribution of radiocesium($^{137,134}$Cs) released from the accident of the Fukushima Daiichi Nuclear Power Plant is mapped. The reconstruction parameters used for each measurement were identical, indicating that the same algorithm can be used for both point or distributed sources.

**INTRODUCTION**
Gamma-ray mapping enables the localization and identification of distributed and point radioactive sources. Fusing data from contextual sensors, such as visual cameras or LiDAR instruments, with data from radiation sensors using Scene Data Fusion (SDF) provides situational awareness in the form of 3-D maps of the area of interest. This information, particularly when obtained in real time can be indispensable for directed search for lost or stolen sources, consequence management after the release of radioactive materials, or contamination avoidance in security-related or emergency response scenarios. In addition, it provides unprecedented means to visualize radiation in 3-D relevant in the communication with the public to address concerns about extent and impact of radiological contamination.

Deploying a 3-D gamma-ray mapping capability on robotics and autonomous systems, such as unmanned ground or aerial vehicles, provides a variety of advantages over human-deployed or manned vehicles. For example, for nuclear decommissioning and environmental remediation efforts, 3-D gamma-ray mapping from a UAS has the potential to improve operational efficiency and streamline decontamination and cleanup efforts, as well as improve operator safety. In case of events leading to changes in the location of radiological materials in so-called legacy or clean-up sites, a fast and accurate assessment will be required to minimize the impact of these changes. At large decommissioning sites, such as the 586-square-mile Hanford site, the UAS could be useful for mapping waste tank storage areas or large fields, buildings, or tunnels around the expansive site more quickly and safely than workers with handheld systems. The UAS also provides a safer way to map large or inaccessible areas with challenging terrain or dangerous conditions without increased human risk or the loss of expensive/large, manned ground vehicles. As the UAS can fly at low altitudes (1-10m) it also enables more detailed mapping of an area of interest than manned helicopters can. The higher precision of 3-D maps over currently available 2-D maps in identifying the location of contamination or hotspots further enables more efficient planning and response. For example, identifying specific hotspots can direct and focus decontamination efforts to smaller areas, thereby potentially reducing cleanup costs.

BACKGROUND

Gamma-ray mapping has been used for geologic survey, contamination mapping, and source search for decades [4,5]. Historically, this method has employed large scintillator crystals, geospatial reference data, and a variety of vehicles to produce geo-referenced maps. Aerial gamma-ray mapping has previously been limited in spatial resolution by flight altitude which directly impacts the system response for so-called proximity sensing. The resulting limit in resolution, coupled with limited data for topographic corrections, reduces the utility of these systems when resolution on the order of meters is desired for mapping. When considering localization accuracy down to meters, these systems have been demonstrated. For example, SDF was performed from a manned helicopter to localize a source on a building in 3-D by creating ground maps from the onboard cameras [11,14]. Recent advancements in UAS technology enable finer resolution aerial gamma-ray mapping, as the UAS can fly closer to objects of interest. Previously, single beam laser rangefinders have been coupled with GPS to facilitate gamma-ray mapping for terrains or surfaces with the goal to achieve resolution on the scale of a few meters [3].

In this work, we adapt the SDF concept [1], a technique that combines 3-D scene mapping capabilities with data from radiation sensors, to enable aerial gamma-ray mapping. Previously, SDF was demonstrated with a coplanar grid CdZnTe-based gamma-ray imaging sensor, the High Efficiency Multimode Imager (HEMI), in a handheld configuration with a Microsoft Kinect in Berkeley, CA and in Fukushima Prefecture, Japan [1,2]. However, to use the Microsoft Kinect, HEMI required an external laptop and tablet computer to process data and produce a 3-D map in real-time. HEMI was also deployed on an unmanned RMAX helicopter in Fukushima Prefecture, Japan to demonstrate the SDF 3-D mapping capability from visual data [13]. Though HEMI was too heavy to be deployed on smaller unmanned aerial systems, the RMAX helicopter has a payload capacity up to 20kg. The HEMI system, in this configuration, required offline processing of collected data.

Here we demonstrate SDF with a compact non-imaging detector array, onboard an UAV and using a multi-beam LiDAR sensor to construct dense 3-D models with centimeter accuracy. These models are used to construct volumetric representations of scenes. This information, along with the system's path through the scene, is combined with data from a 2 x 2 array of CsI(Tl) detectors to enable real-time meter resolution gamma-ray mapping in both unmanned aerial and handheld measurements. This is different from past approaches in that this capability is robust in both indoor and outdoor environments, can construct 3-D volumetric scenes, demonstrates 3-D self-shielded proximity mapping with commercial radiation detectors, and computes all data products onboard the LAMP system with real-time telemetry from the UAV.

**LOCALIZATION AND MAPPING PLATFORM (LAMP)**
The Localization and Mapping Platform (LAMP) is a compact system developed to enable modular, customizable configuration of different contextual and radiation sensors for specific use cases. LAMP is platform-agnostic and self-sufficient, meaning it can be deployed in handheld configurations or on unmanned ground or aerial vehicles, and in multiple configurations for different application areas without external power or offline processing. Reconstruction parameters are not changed between configurations, so the appropriate vehicle can be chosen for the measurement scenario.

The left image in Fig. 1 shows a model of the LAMP prototype configured for deployment on an UAS. Key hardware components include an Intel i7 computer, power distribution system, Velodyne Puck Lite LiDAR [8], Vectornav 300 [10] inertial navigation system, visual camera, and 131.1 cm$^3$ (8 inch$^3$) of CsI(Tl) scintillator volume. In the results shown here, the GPS and camera data were not used. The weight distribution of these components is described on the right of Fig. 1. As this figure shows, the LAMP system weight is composed of more than 25% detector components, which is important as most UAV systems have dramatic tradeoffs with payload weight and flight time (e.g. as payload weight increases, flight time decreases) [6].

This UAS configuration of LAMP has a total mass of 4.5kg, which includes the sensors shown on the left of Fig. 1. For this weight, LAMP has a battery life of about 1 hour at full-load computation and data collection time a. When no computation is being done on the system, the battery life can sustain 3.5 hours of data collection. For demonstration measurements, LAMP was deployed on a DJI Matrice 600 UAV [6], which has a maximum payload of 6kg, as shown in Fig. 2. In the current configuration, a flight time of about 20 minutes is possible.

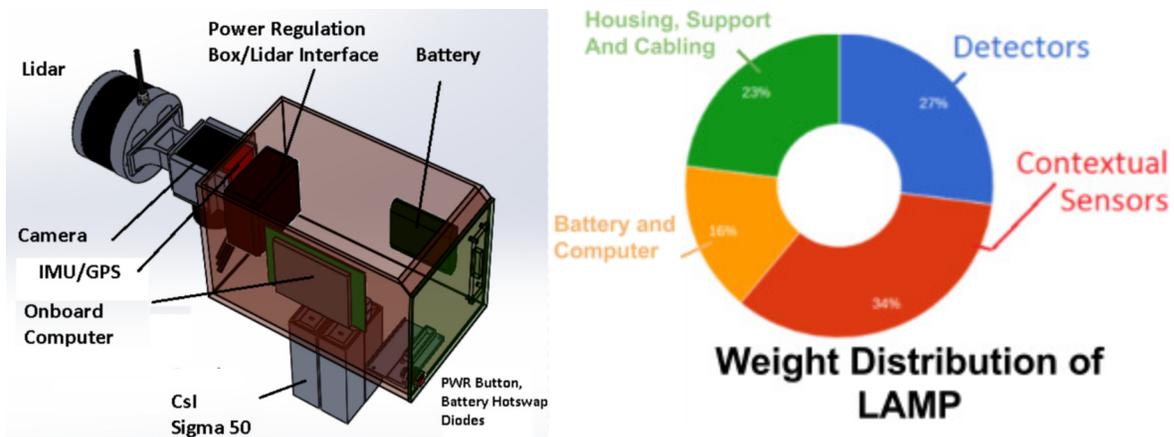

Figure 1 (Left): The LAMP with a suite of sensors that includes a VectorNav300 INS, Velodyne Puck Lite LiDAR, onboard computer and ~590g of CsI(Tl) scintillator (four Kromek Sigma50 sensors) [9]. (Right): Weight distribution for LAMP system - the sensor mass is around 65% of the total system mass.

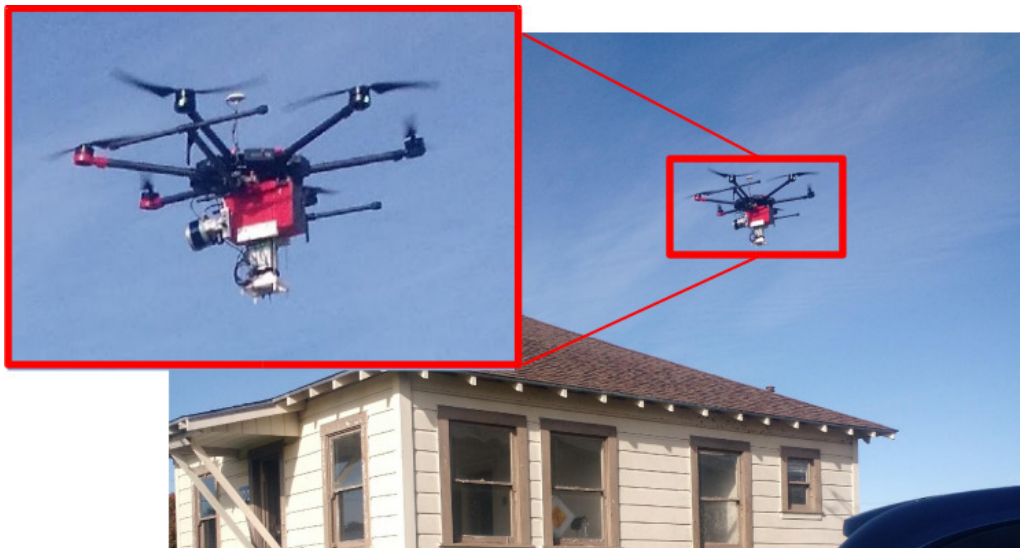

Figure 2: LAMP deployed on a DJI Matrice 600 UAV at a University of California, Berkeley (UC Berkeley) campus test site.

**3-D PROXIMITY GAMMA-RAY MAPPING METHODOLOGY**

Gamma-ray SDF mapping fundamentally relies on the 3-D scene data position and orientation estimates provided by contextual sensors. Simultaneous Localization and Mapping (SLAM), achieved using Google Cartographer [12], enables LAMP to estimate positions with centimeter accuracy and orientation in real-time. LAMP can then apply SDF concepts to signal modulation from the path through an environment, with the 2 x 2 array of commercial radiation sensors. A sparse voxel grid is computed from the model points which compose the volumetric scene. MLEM [7] is computed on the volumetric space from the radiation data in order to estimate the volumetric distribution of a source, which is shown schematically in Fig. 3. This approach estimates the source distribution in the scene and makes no assumptions about the source extent. We apply a GPU-accelerated 3-D real-time voxelized MLEM algorithm to this data to estimate this distribution. The MLEM algorithm requires the computation of both the system matrix, as well as an estimate of the sensitivity which describes the relationship between the measurement path and the 3-D measurement environment.The system matrix is continually recomputed as the path and 3-D model are updated throughout the duration of the measurement. The effects that are taken into account in the system matrix include inverse squared source strength fall off and the directional response (cf. next section) of each of the four CsI(Tl) detectors.

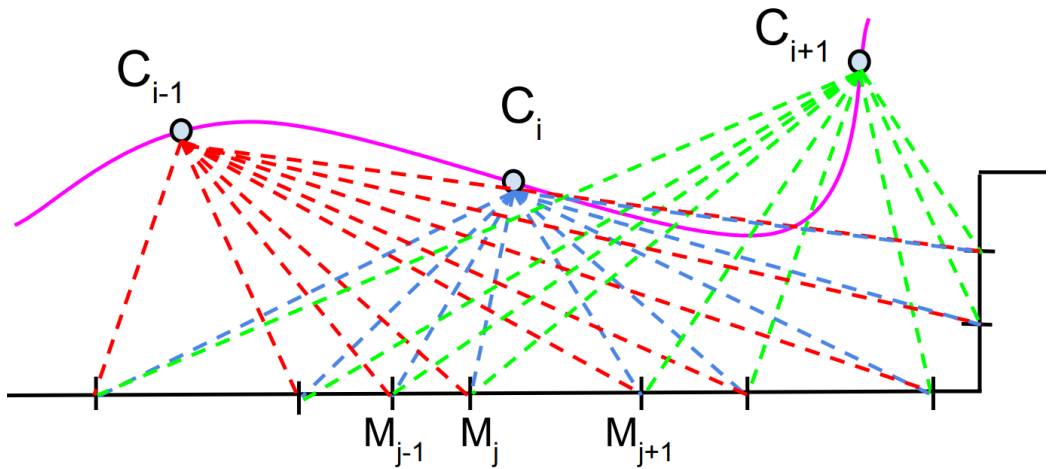

Figure 3: A schematic of the LAMP MLEM problem, where the SLAM solution already exists. Here $C_i$ represents the radiation data from each position and $M_j$ represents the model points derived from SLAM as points or 3D voxels.

**SYSTEM ANGULAR RESPONSE**

Each radiation detector in the 2 x 2 detector array has a unique angular response due to partial attenuation of radiation signal through neighboring detectors. This effect, referred to as self-shielding, can be measured and is used as part of the MLEM computation performed on LAMP to create the 3-D reconstruction. Generation of the angular response can be performed in the flight configuration, where LAMP uses SLAM to track its motion through a scene relative to a known source position, as shown in Fig 4. In this case, the UAS was flown around the point source and rotated 360 degrees at various angles relative to the source. Pose estimates from SLAM are used to compute the distance and orientation from a point source, which enables the computation of the system angular response. Fig. 5 shows the response functions for the four CsI(Tl) detectors. These response functions assume symmetry about zero in the polar angle, as only the lower hemisphere response was measured in this example.

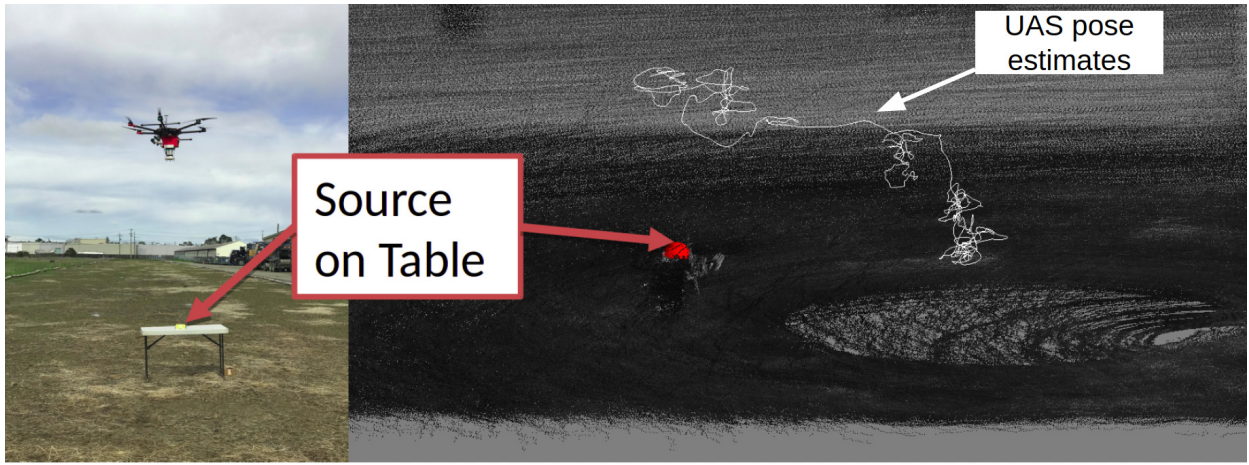

Figure 4: Characterization of the angular response of the radiation detectors on LAMP from an UAS flight with known source location. (Left) A photograph of LAMP flying on the UAS over a table where a $^{133}$Ba source rests. (Right) The point cloud model of the table and flight path. LAMP can characterize the system response from identification of the radiation source placard in the point cloud and knowledge of the system pose. The latter of these is provided directly from SLAM. The figure shown does not represent all data used to compute the system response provided in Fig. 5.

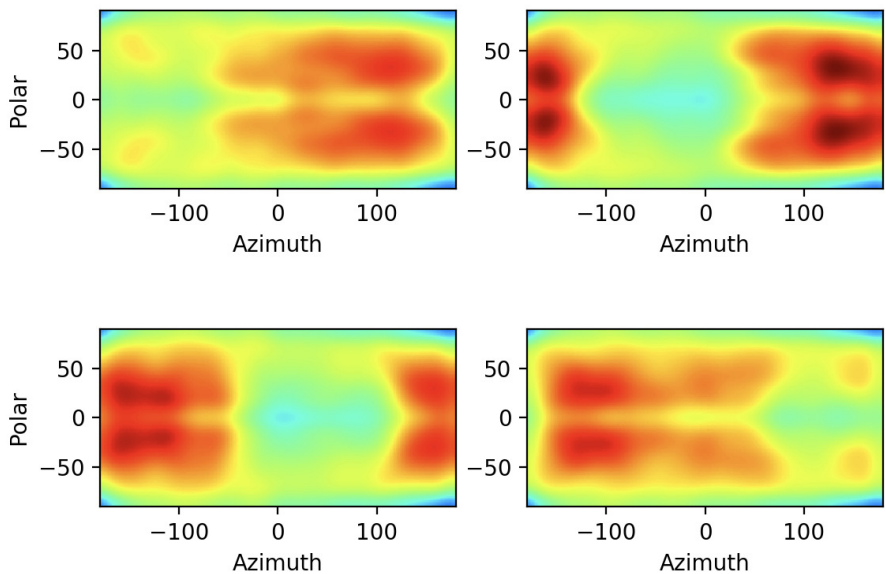

Figure 5: Result of the system characterization measurement shown in Fig. 4. The source response for each of the 4 detectors is shown, where the 2 x 2 array partially attenuates the source based on the orientation and elevation of the system relative to the $^{133}$Ba source.

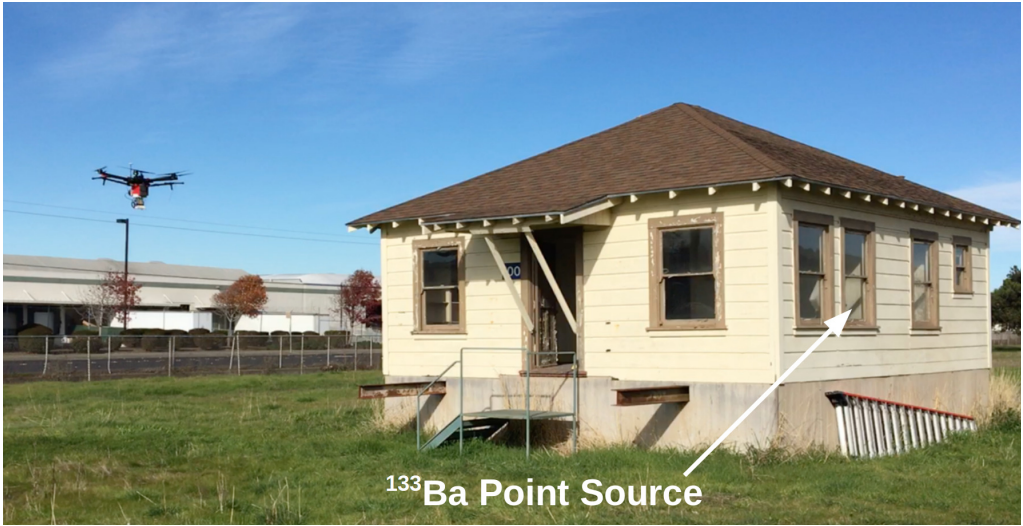

Figure 6: Unoccupied test building at the University of California, Berkeley. A 400uCi $^{133}$Ba source is placed just inside one of the windows.

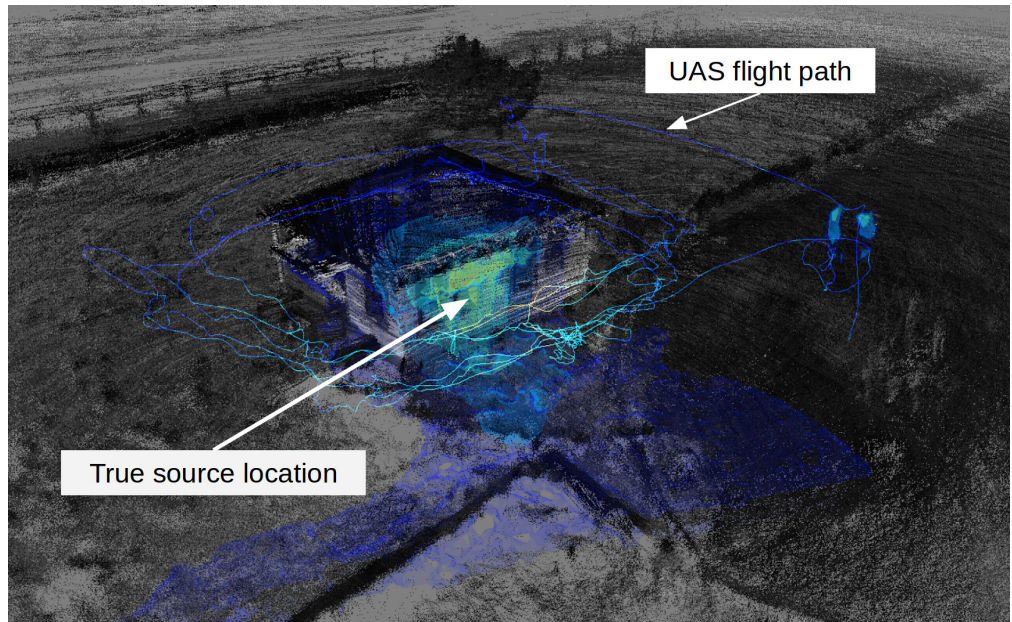

Figure 7: MLEM result from a 400μCi $^{133}$Ba point source localization measurement. LAMP was flown on a UAV around the test building shown in Fig. 6. The total flight time of this measurement was 6 min. The dark blue to light blue colors along the UAS flight path around the house represents the count rate of the detectors as the system took-off, flew around the house, and landed. The dark blue to light green contours show the MLEM result. The house dimensions are approximately 5m x 5m at the foundation. The total extent of the point cloud model representing the environment is approximately 50m x 100m.

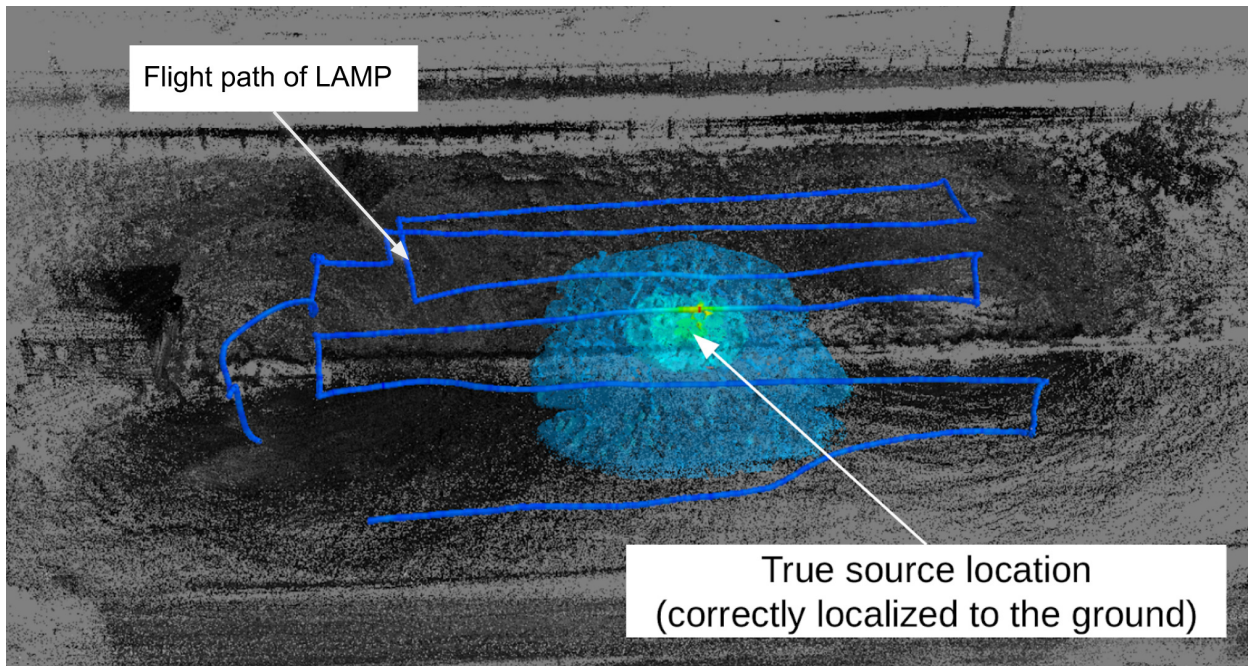

Figure 8: MLEM result from a different localization measurement for a 400μCi $^{133}$Ba point source in the middle of a field at a UC Berkeley test site. The flight path is shown over top the MLEM result, with a relative intensity color scale. A constant height raster was performed at an altitude of about 2.5m. The area mapped is about 30m x 20m, in about 8min.

**RESULTS**

A possible source search scenario was simulated at a University of California, Berkeley test site where a 400uCi $^{133}$Ba source was placed inside of an unoccupied test building in the middle of a field. A survey of the building was conducted by flying LAMP on a DJI Matrice 600, as seen in Fig. 7 where the flight path is shown in blue. The flight and measurement lasted about 6 minutes and the source was localized within the house. The true source location was inside one of the windows, as was indicated in Fig. 6. Three iterations of MLEM were performed on the voxelized 3-D space.

A separate source search scenario was demonstrated as shown in Fig., 8, during which LAMP mapped an area of approximately 20m x 30m, the measurement time was about 8min. The flight height was about 2.5m above the field and the 400uCi $^{133}$Ba source was localized with MLEM with three iterations. While the localization of the source is very accurate, the reconstructed source extent is much broader than the physical source. Here the reconstructed source extent would depend on the distance of closest approach and the number of MLEM iterations.

There are many factors which contribute to the broadness of this distribution which are unrelated to the physical source (the source is several millimeters in its largest dimension). The largest limiting factor is that the system relies on the combination of proximity imaging with self-shielding of the four detectors. Because the angular resolution of the system is so poor, we primarily rely on the proximity effect for localization. Localization by proximity is essentially limited to the distance of closest approach to the source. In the reported cases, LAMP's distance of closest approach is a 1-3m, leading to several meter resolution. Additional effects that limit resolution include: 1) the down-scattered photon flux is not accounted for in this image, 2) attenuation of source intensity in the scene is not taken into account and 3) geometric symmetries that imply uniform angular response, especially directly below the 2 x 2 array. The first of these could be remedied with photo-peak windowing techniques, with an associated efficiency loss. The third effect, due to the symmetry below the 2x2 detector array, is related to the limited signal modulation from sources directly below LAMP, when attached to the UAV. This lack of modulation reduces the localization ability, projecting intensity onto the ground. This effect leads to the artifact seen in Fig. 5 where there is increased intensity below the path of the system.

**DISTRIBUTED SOURCE MAPPING DEMONSTRATION**

In March 2017, LAMP was operated in handheld mode and walked through a parking lot in the Okuma city, Fukushima Prefecture, Japan. One result demonstrating distributed source mapping is shown in Fig. 9. This area was located within the exclusion zone in Fukushima Prefecture at the time of the measurement, where $^{137,134}$Cs is the primary contaminant. The measurement area consisted of a parking lot with vegetation growing in the periphery and within cracks in the pavement. The path traversed through the scene, colored in relative intensity for count rate, and the MLEM result are shown in relative intensity. This scene is interesting because several of the physical features correlate with the gamma-ray data. The first of these features is that the asphalt surface of the parking lot does not facilitate the accumulation of radioisotopes, while areas with accumulated vegetation, corresponding to cracks in the asphalt or areas of lower elevation, tend to facilitate the aggregation of radiocesium. This phenomenon is observed at the periphery of the measurement area where the primary hotspots were localized to a water drainage area, where cesium could accumulate. Additionally, shown in Fig. 9 are lines of vegetation that have grown through the cracks in the pavement and are sometimes correlated in intensity, in the count rate and source distribution shown in the MLEM result. The hotspot shown in Fig. 9 also correlates with the vegetation at the back of the lot. This measurement was performed in approximately 4.5 minutes walking through the parking lot once.

The scene from Fig. 9 is complex and, in general, distributed source environments are difficult to provide ground truth for, as traditional measurement techniques produce limited data about the scene or are labor-intensive. Correlation of the physical features, dose-rate, and traditional gamma-ray spectrometry are the current proxy for verification. Future work in quantitation, the correspondence of MLEM intensity to total activity, is under development.

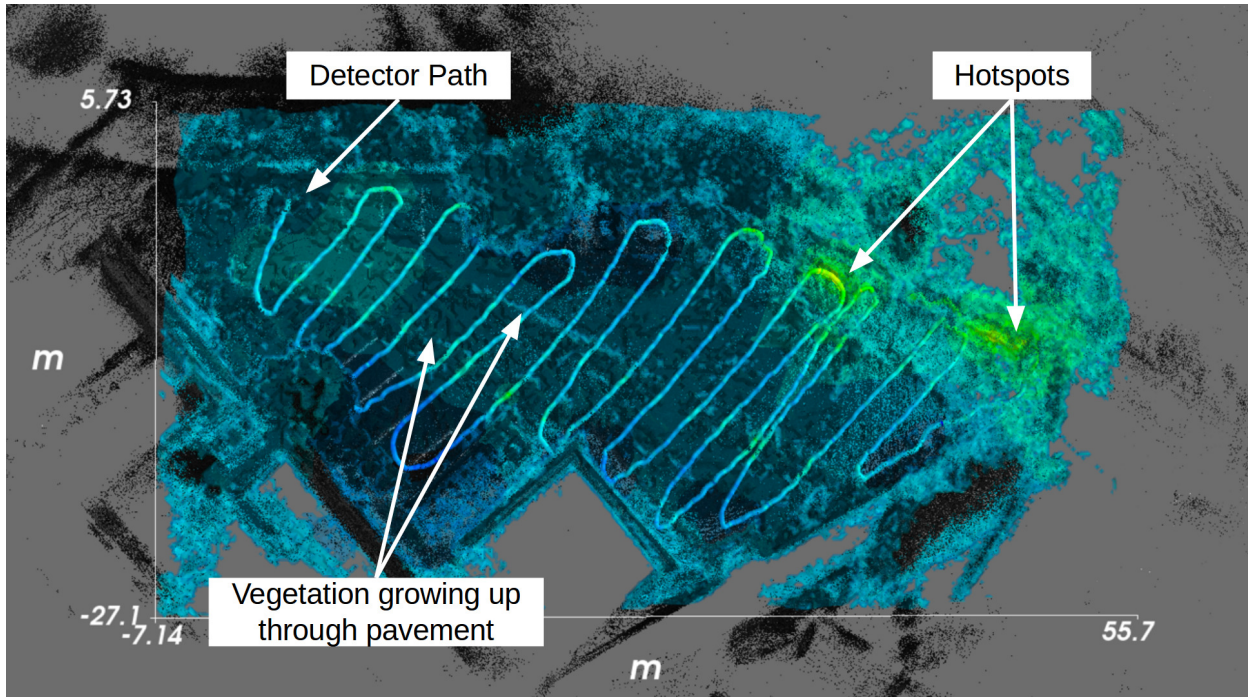

Figure 9: Contamination mapping inside the Fukushima Exclusion Zone via handheld LAMP measurement. The system was held about one meter above the surface in which this measurement occurred. In this environment, the contamination is mostly $^{137,134}$Cs from the Fukushima Dai'ichi Nuclear Power Plant accident. A top-down view of the lot is visualized, with vegetation noted at the periphery and growing in cracks in the pavement. Hotspots were localized in vegetation at the periphery of the parking lot by a water drainage area. All data shown was collected in about 4.5min. Dose rates in this area were, on average, about 4 uSv/hr.

CONCLUSION

In this paper, we have introduced LAMP as a modular, robust, indoor and outdoor measurement platform capable of being deployed for UAS measurements. We have demonstrated LAMP and the SDF concept as a gamma-ray mapping capability for a simple source localization case and for complex distributed source mapping. The current LAMP detector payload enables proximity-based, volumetric gamma-ray mapping via SDF with a compact array of non-imaging detectors combined with SLAM. We have discussed the limitations of this method, as well as the source of these limitations. LAMP represents a substantial improvement to conventional non-imaging radiation detectors by demonstrating the ability to associate 3-D objects with gamma-ray signatures to produce data-rich gamma-ray maps.

LAMP provides the basis of fast integration of many different contextual sensors and enables 3D gamma-ray mapping even with simple, non-imaging radiation detectors. Since cost is a large factor in any radiation detection system, integration of commercial sensors provides a way to reduce costs associated with real-time 3D radiation mapping. These benefits are provided by the inverse-square law of proximity mapping, and the spatial integration that MLEM or backprojection portray.

ACKNOWLEDGEMENTS

This material is based upon work supported by the Defense Threat Reduction Agency under IAA numbers 10027-21370 and 10027-23334. This support does not constitute an express or implied endorsement on the part of the United States Government. In addition, the Fukushima measurement is based upon work supported by the Japan Atomic Energy Agency under WFO contract number FP00002328.